\documentclass{ws-mpla}
\usepackage{graphicx}
\begin{document}

\markboth{J. E. Horvath and G. Lugones} {Another decade of strange
quark matter (astro) physics}

%
%

\title{ANOTHER DECADE OF STRANGE QUARK MATTER (ASTRO) PHYSICS}

\author{\footnotesize J. E. HORVATH$^*$ and  G. LUGONES$^{\dag}$ }

\address{Instituto de Astronomia, Geof\'\i sica e Ciencias Atmosf\'ericas\\ Rua do
Mat\~ao 1226, 05508-900 S\~ao Paulo SP, Brazil\\
$^*$foton@astro.iag.usp.br\\ $^{\dag}$ glugones@astro.iag.usp.br}

\maketitle

\pub{Received (Day Month Year)}{Revised (Day Month Year)}

\abstract{We present a perspective of strange quark matter research
in the 1991-2001 decade focused on astrophysics topics, with
particular attention to open problems. We outline the basic
concepts and developments in the field, paying attention to the established
research and promising perspectives. An analysis of the whole literature
(experimental searches, theory,
astrophysical/cosmological) serves to point out some general trends,
supported by a fairy complete statistical sample of published works. }

\section{Introduction}

\bigskip

The history of sciences register many cases of (almost)
simultaneous works giving rise to an entirely new area of
research. It is also quite common to be able to track backwards
the branching of an existing area, generally much after those
initial events. A less common situation is to witness that process
from the start, and additionally being able to collect virtually
all the papers which in turn help to, identify the main trends.
Strange quark matter (SQM) belongs to this interesting category.
Because of the quick development of several ideas about SQM, a
benchmark event (Workshop on SQM in Physics and Astrophysics,
Aarhus University, Denmark 1991) \cite{AA}  made possible to join
a large fraction of the active researchers in the field. Even
today, and taking into account a considerable growth in the number
of the latter, it would be possible to repeat the achievement (a
few conferences have covered a major portion of SQM field in the
following years, although other current topics were a main part of
the program). Actually most of the colleagues that contributed to
SQM physics/astrophysics are alive and still active, while several
others enter the field each year.

In addition, and as a case study for the history of sciences, SQM
presents some peculiarities worth noting. Perhaps the most
notorious is that after decades of existence it is still
completely speculative, yet vigorously pursued in several
directions. How did this situation affected the form in which the
research is done? Which are the main avenues streets and
sidewalks? The purpose of this review is not to solve these
problems, but rather to present an assessment of assorted
astrophysics, with the aim of formulating quite clearly a few
persistent questions about SQM which often remain hidden or
implicit. We have no intention of writing a full review (many
excellent ones have appeared over the years, see the reference
list), but rather to offer a personal perspective of the evolution
we have witnessed and future trends. In the present work we will
focus on the last decade (1991-2001) astrophysical developments
mainly. Some facts will be substantiated by a brief (but almost
complete) compilations of the bibliography of the period,
available at the address www.iagusp.usp.br$\setminus\sim$mpallen
in .txt form which may be considered as the electronic companion
to the Aarhus compilation.

\section{What is exactly strange quark matter? }

The concept of elementary constituents of nucleons (quarks and gluons) is
clearly central to SQM and preexists it, therefore it is useful to repeat
some historical developments for completeness. With increasing center-of-mass energy,
experimental searches of the elementary
components (partons) of protons and other hadrons revealed a whole new realm
of physics between the second half of the '60s and the early '70s. The search of a
theoretical framework to engulf this body of knowledge was developed in parallel, first
focused on classification schemes (or, as is called today, flavor physics) and later on
finding a theory to describe the dynamics. The strong motivation given by
developments of gauge theories
in the '70s eventually rendered the non-abelian version based on the
$SU(3)_c$  symmetry group \cite{QCD} as a natural candidate for a theory of
strong interactions. The new quantum
number carried by the elementary constituents (quarks) was dubbed "color",
and thus the dynamics involving quarks and gauge fields (gluons) become
known as Quantum Chromodynamics (or QCD for short).

On the experimental side, repeated efforts to find these entities
as free particles (asymptotic states) failed, and convinced most
people to accept a striking feature of the theory: that the
interactions preclude the appearance of the quarks and gluons
outside hadrons, they are instead confined inside them. Moreover,
another property was soon demonstrated to liberate quarks and
gluons but only for momentum transfer scales $Q^{2}$ large enough.
This is the so-called {\it asymptotic freedom}, and states that
the colored particles behave as if they were free in the limit
$Q^{2} \, \rightarrow \, 0$. There is an energy (or momentum)
scale above which color  quantum number is not confined any more,
but how large the momentum transfer should be (or in other words,
which is energy, as measured by the temperature or density of the
ensemble allowing the deconfinement) is still a matter of
controversy. There is no doubt that the early universe passed
through a deconfinement $\rightarrow$ confinement phase transition
along its cooling, but there is less certainty that the densities
of the "natural" laboratories (neutron stars) in which compression
would deconfine hadronic matter are high enough to do so. In fact,
the earliest calculations \cite{Baym} using reasonable models for
both the confined and deconfined phases imprinted on successive
researchers the uncertain conclusion that quarks and gluons
(forming a state known as the quark-gluon plasma, or QGP) should
appear at densities above, say, $10 \times \rho_{0}$; with
$\rho_{0}$ the nuclear saturation density.

The latter statement is an example of the uncertainties and types
of loopholes which plagued the attempts to determine the
transition points, and also the nature of the transition itself
(at least when full numerical calculations \cite{num} were out of
sight ). If carefully scrutinized, most of the times the
conclusions are extracted from simultaneous extrapolations of both
a quark model, expected to be valid for $\rho \rightarrow \infty$,
and an hadronic model valid around $\rho_{0}$ but uncertain much
above it. There is no certainty in either one and hence in the
final result, not to talk about the "induction" of a definite
order of the transition because of the adopted functional forms of
the thermodynamical quantities of both sides. Nevertheless these
serious and honest attempts have proliferated until today, given
that the transition is still elusive (the better studied finite
temperature case still has some small uncertainty in the value of
$T_{c}$ and a better assessment of the order, see Ref.5 for
details).

Independently of the above caveats, a much radical proposal emerged to
complicate even more these matters. The asymptotic freedom property guarantees that
quarks and gluons will be the ground state of QCD at high densities/temperatures, but
says nothing about the ground state at lower densities or temperatures. The everyday
experience strongly suggests that ordinary hadrons confine the quarks/gluons and thus
constitute the "true" (in the sense of $\rho \rightarrow 0$ and $T \rightarrow 0$)
ground state of hadronic matter. The strange matter hypothesis comes precisely to
challenge this "common sense" statement: it says that the true ground  state of hadronic
matter is a particular form of the QGP , differing from the
ordinary matter by the presence of a key quantum number (strangeness). This
is counterintuitive to many people, but a careful look at the physical
arguments shows no inconsistency whatsoever, at least in principle.

The argument for the SQM being the true ground state goes as follows: as is
well-known the quantity that determines which phase is preferred is the Gibbs free
energy per particle $G/n$ as a function of the
pressure ( we impose $T=0$ hereafter as appropriate for highly
degenerate hadronic matter, it is easy to see that the term $-TS$ in the
free energy disfavors SQM at high temperatures). As $P$ is increased starting from the
neighborhood of the nuclear matter saturation point $\rho_{0}$ the
asymptotic freedom says
that there has to be a switch from nuclear matter (N) to elementary hadronic
constituents, that is, the lighter quarks $u$ and $d$. The point at which this is
supposed to happen has been labelled as "$P_{c}$" on the horizontal axis of Fig. 1
(therefore, the
doubts stated above about the appearance of the QGP inside neutron stars may be
now restated as whether the pressure at the center is larger or smaller than $P_{c}$).

\begin{figure}
\includegraphics[angle=-90,width=12cm,clip]{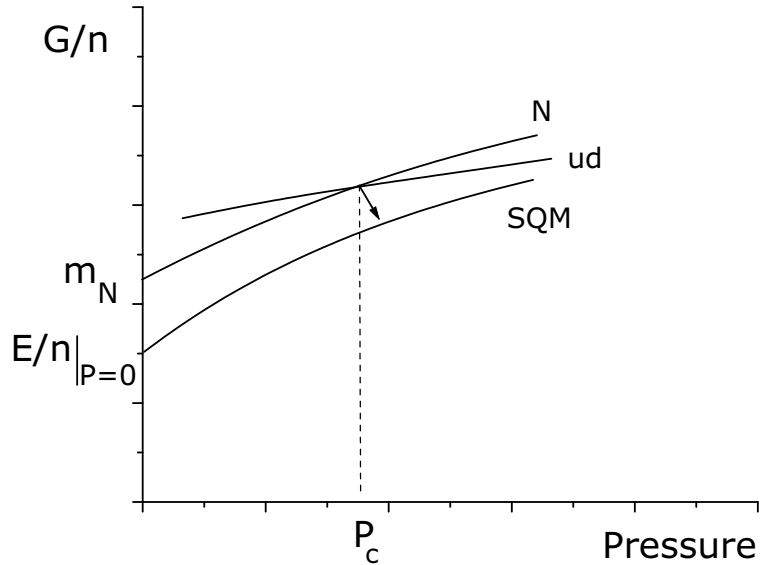}
\caption{The energetics of SQM. Gibbs free energy per particle
$G/n$ is shown for neutron $N$, two-flavor quark matter $ud$ and
strange quark matter $SQM$. At $P = 0$ the free energy per
particle of $SQM$  is (by hypothesis) below the mass of the
neutron $m_{N}$. Although the $SQM$ curve is always below both,
its production is possible only after deconfinement of the $ud$
matter, being suppressed by the strangeness content below $P_{c}$.
The decay of $ud$ matter to $SQM$ is a transient depicted with an
arrow and starts somewhere beyond $P_{c}$ depending on the
nucleation conditions.}
\end{figure}

However, it is here where the concept  of strangeness comes in.
Strangeness is the flavor quantum number carried by $\Lambda$'s
and other heavy hadrons. At the elementary level, it is carried by
a different quark $s$. While creating strangeness in hadrons costs
energy (because strange hadrons are heavier than non-strange ones;
for instance, $\Lambda$'s are heavier than neutrons and so on);
this is reversed inside the QGP. The reason is simply the Pauli
exclusion principle: a new Fermi sea in the liquid (the one of the
$s$ quark) allows a rearrangement of the energy, and this sharing
lowers the energy per particle. How much the gain is is not
precisely known, but it is not impossible to lower the free energy
per particle to a value that would be lower than the mass of the
neutron $m_{n}$ even when $P \rightarrow 0$. If realized, this
would preclude the (strange) QGP to decay into ordinary hadrons
because this would {\it cost} energy and the SQM would have been
created. Put it simply, the compression would liberate the
elementary components that quickly create its own way of
surviving.  We stress that all these are bulk (i.e. large number)
concepts, and it is central to the SQM hypothesis to reach a
strangeness per baryon of the order one (and exactly one if the
strange quark had no mass to deplete its relative abundance). This
is not possible in a few body system like a nucleus, because each
weak decay creating a strangeness unit contributes roughly with a
factor $G_{Fermi}^{2}$ to the amplitude, and thus the simultaneous
decays are strongly suppressed; this is why it has been very
difficult to produce even doubly strange nuclei, let alone higher
multiplicity ones. However, once quarks roam free in the QGP they
can easily decay by $u + d \, \rightarrow \, u + s$ because there
is plenty of phase space for the products until equilibrium is
reached (see Fig. 2). The bulk picture has been always one way or
another behind the idea of SQM.

\begin{figure}
\includegraphics[angle=-90,width=12cm,clip]{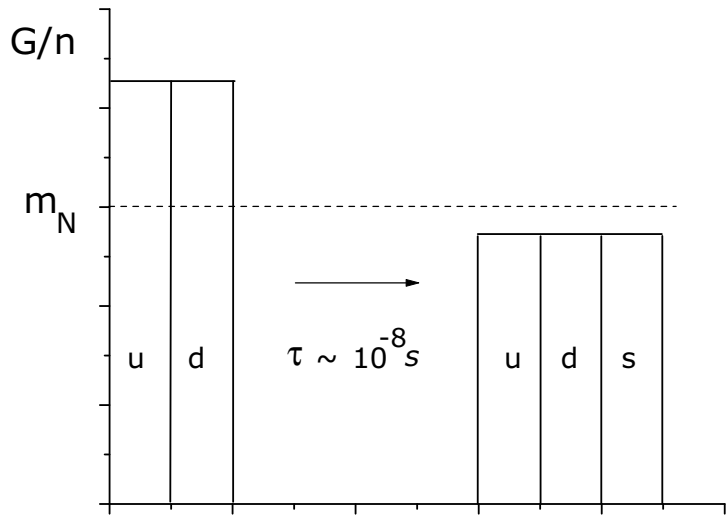}
\caption{Another view of the SQM hypothesis. The Fermi seas of $u$
and $d$ quarks lie above the neutron mass $m_{N}$ initially.
Because creating net strangeness lowers the energy per baryon,
weak interactions will do it in a timescale of $\sim \, 10^{-8} \,
s$. The energy per baryon is lower now, but the actual question
related to SQM is whether $G/n$ stays above $m_{N}$ or drops below
it, as shown in this graph. If the latter situation happens in
dense deconfined matter, SQM would be the ground state of hadronic
matter.}
\end{figure}

As it stands, the SQM hypothesis is very bold. It says that
everything we see around us is in a metastable state, and if
conditions for creation of a large net strangeness were met, the
matter would not make back ordinary hadrons (technically it is
said that SQM constitutes a non-topological soliton stabilized
against decays by a conserved charge, the baryon number, see Ref.
6 for a thorough discussion of this case and related ones). The
general idea of reaching extreme conditions and stabilizing the
QGP is already apparent in the paper of Bodmer \cite{Bodmer},
later reintroduced and refined in references \cite{CK,Tera,Bjo}
and colorfully discussed in the paper of Witten \cite{Ed}, which
was fundamental to give a big boost to SQM research.

Many applications of SQM in astrophysics were foreseen during  the
first decade after its official birth \cite{Ed} and early infancy
\cite{FJ}. Since that the field has broadened a lot
and a variety of problems are being studied presently.
Nevertheless, key questions of SQM such as whether it does exist
or not, and whether it has been ever produced in the Universe are
still unsolved. On the other hand, we begin for the first time
to have the possibility of falsifying these basic questions mainly
thanks to the new generation of space telescopes (HST, Chandra, XMM)
and neutrino observatories (SNO, Kamioka, Icecube),
to name just a few.

Since astrophysical insight has shown to be essential for the
study of fundamental questions related to SQM, we shall focus
briefly in a few selected astrophysical problems, trying to give
an assessment and pointing on the uncertainties and possible
advances that may be expected in the near future.

\section{SQM in supernovae}

As a ''natural'' environment in which SQM might form, some authors
have advocated  for core collapse supernovae. The reasons to
expect an important role of SQM there are many. First, despite of
more than three decades of theoretical research and hard numerical
modelling, the processes that cause the explosion of massive stars
are still not understood \cite{Bur}. If, as the more recent and
detailed numerical simulations suggest, the neutrino-driven
mechanism is fundamentally flawed, the current paradigm for
explaining massive star explosions would have to be deeply
revised. Although it is still too early for making definitive
conclusions, investigations including the possible transition to
deconfined QCD phases gain potential importance. The first studies
of SQM in supernovae \cite{Mac,Trio,LBV,DPL,Anand} showed that
this hypothetic subnuclear energy source is more than adequate to
contribute to the explosion, and that some observed
characteristics in the neutrino emission of SN1987A may be
naturally explained within this scenario \cite{Hat,SatSuz,Mac}. It
is worthwhile to note that a second peak in the neutrino emission
is {\it predicted} in these models, and such signal has been
tentatively associated to the late neutrinos from SN1987A detected
by Kamiokande (which have to be otherwise interpreted as a
statistical gap within the current paradigm). As long-term
forecast, supernovae are perhaps the only astrophysical events in
which we could have the possibility of making a
''multiwavelength'' detection (neutrinos, various electromagnetic
wavelengths, gravitational waves) of the process of SQM formation.
An accurate theoretical knowledge of the process of formation
could allow us to presence it in real time, in case it actually
occurs. However, this approach is still in its infancy, and just
bold expectations have been formulated. In addition, a firmer
observational background would be needed, which needs nothing more
than the occurrence of a number of supernova explosions in the
neighborhood of our galaxy. Second, although the general picture
of SQM formation in supernovae has been qualitatively constructed,
no systematic calculations have been made. There are also many
unresolved questions related to strong interactions at high
densities, which introduce an uncomfortable degree of uncertainty
in all conclusions.

A ''direct'' mechanism for SNII explosion driven by SQM formation
has been in fact advocated in the original papers by Benvenuto,
Horvath and Vucetich \cite{Mac,Trio}. The basics are as follows.
As discussed and agreed in the literature, a seed of SQM must
become active or form following the standard bounce. The interface
must then propagate outwards powered by the energy release of
converted neutrons, much in the same way as a laboratory
combustion. It seems reasonable to assume the combustion to begin
as a laminar deflagration, which quickly reaches a regime of
turbulent deflagration \cite{HB,Lug}. It is still not clear
whether the detonation mode is feasible (as originally suggested),
since it requires fast transport of heat to sustain the front.
Assuming the latter case, and since the conversion is not expected
to be exothermic all the way down to zero pressure, it is
unavoidable that a detonation will become a standard shock wave
beyond some radius (within the MIT Bag model for SQM this radius
is the one for which $E - 3P= 4B$). This shock wave will propagate
outwards and the question is whether or not it will be able to
transfer its energy and complete the work unfinished by the
unsuccessful prompt shock wave. A less extreme combustion mode
(subsonic but still very fast) may be the final outcome instead of
a detonation, and its propagation would mix the material on
macroscopic scales due to the action of Landau-Darreius and
Rayleigh-Taylor instabilities, but its role in the reenergization
of the stalled shock has not been calculated as yet.

If energy can not be directly transferred to the outer layers, SQM
formation may still be important because of the production of neutrinos
by appropriate reactions in the deconfined phase. The binding energy
of the strange star has to be released as well \cite{BomDat} , much in the
same way as the binding energy of the neutron star in the standard picture.
Although new fresh neutrinos could in principle  produce a late
revival of the stalled shock wave, other features than the total
released energy are essential such as spectral features of
the neutrino emission, and more importantly (if the transition
happens to be somewhat delayed) the exact time of its occurrence, since if
it occurs too late there will be no way to explode the star by the
shock reheating mechanism at all.

A better understanding of the previous sequence of combustion
processes will also give information about the timescale of the
conversion of the star, which is closely related to the different
observational signals. These calculations are yet to be performed
in detail and constitute a priority task for the near future.

\section{Delayed conversions, compact star structure and gamma-ray bursts}

If the just-born protoneutron star (PSN) does not collapse to a
black hole due to accretion in the early stages \cite{LatPra}, and
it happens that SQM is absolutely stable (i.e. the true ground
state at zero pressure) then pure strange stars, made up entirely
of strange quark matter from the center to the surface, may be the
compact remnants of supernovae. In the case of absolute stability,
if the transition is {\it not} triggered during the supernova
explosion, all ''normal'' neutron stars would be in a metastable
state, which is quite difficult to imagine because of ISM
contamination arguments \cite{AFO,Jes,MTH} and the mismatch
$\tau_{conv} \, \ll \, \tau_{star}$ between the timescale in which
favorable conditions for conversion occurs $\tau_{conv}$ and the
lifetime of the star $\tau_{star}$. According to recent
calculations the deconfinement  transition is more likely to occur
by heating and compression during the Kelvin-Helmholtz phase of
proto-neutron star (PSN) evolution (see, for instance,
Ref.~\refcite{LuBe}). If it did not happen there, once the PNS has
cooled to temperatures below $\sim \, 1 MeV$, only accretion from
a companion star or strangelet contamination would allow the
transition (and many barriers may preclude its occurrence), even
in the case where it is energetically favored. Thus, the existence
of strange stars is determined not only by fundamental questions
concerning the true ground state of dense matter but also by the
exact physical conditions in the specific astrophysical
environments together with the plausibility of the conversion
mechanisms in these situations.

Other  interesting questions are related with the emission of
gamma-ray bursts associated with the conversion process. Many
works in the past have explored the idea that the conversion of NM
into SQM in NSs may be an energy source for GRBs
\cite{AFO1,MX,Hae,CD,BomDat,Chi,Rach}. These models addressed
spherically symmetric conversions of the whole NS rendering
isotropic gamma emission. These models are still very schematic to
address the more difficult questions, and many of them tend to
ignore, for instance, the so-called baryon load problem, briefly
stated as the smallness of baryons in the ejected flux as a
precondition to avoid degradation of the energy to softer bands.
Accumulating observational evidence suggests that at least
''long'' GRBs are strongly asymmetric, jet-like outflows, a
feature that needs some crucial ingredient in the SQM physics
formation/propagation to proceed. To be sure, the ''short'' burst
subclass is not obviously asymmetric, and they may actually be
spherically symmetric if the sources are close enough.

A new potentially important feature recently recognized \cite{Lug}
is that if a conversion to SQM actually begins near the center of
an NS, the presence of a moderate magnetic field B ($ \sim
10^{13}$ G) will originate a prompt {\it asymmetric} gamma
emission, which may be observed as a short, beamed GRB after the
recovery of a fraction of the neutrino energy via $\nu {\bar\nu}
\rightarrow e^{+}e{-} \rightarrow \gamma \gamma$. The basic
physical effect is that the influence of the magnetic field
expected to be present in NS interiors quenches the growth of the
hydrodynamic instabilities in the equatorial direction of the star
(parallel to the magnetic field) while it allows them to grow in
the polar one. As a result, the flame will propagate much faster
in the polar direction, and this will result in a strong
(transitory) asymmetry in the geometry of the just formed core of
hot SQM, which will resemble a cylinder orientated in the
direction of the magnetic poles of the NS. While it lasts, this
geometrical asymmetry gives rise to a bipolar emission of the
thermal neutrino-antineutrino pairs produced in the process of SQM
formation. This is because almost all the thermal neutrinos
generated in the process of SQM formation will be emitted in a
free streaming regime through the polar cap surface, and not in
other directions due to the opacity of the matter surrounding the
cylinder. The neutrino-antineutrino pairs annihilate into
electron-positron pairs just above the polar caps of the NS,
giving rise to a relativistic fireball, thus providing a suitable
form of energy transport and conversion to gamma-emission that may
be associated to short gamma-ray bursts. A unifying scheme in
which SQM appearance produces spherical ejection phenomena to
highly asymmetric gamma beaming, as a more or less continuous
function of the magnetic field $B$ and the astrophysical system
under examination may be possible, and is tentatively sketched in
Table 1.

\begin{table}[h]
\tbl{Possible outcomes of SQM burning in stellar systems}
{\begin{tabular}{@{}cccc@{}} \toprule
Mag. field (G) & Type II SN & LMXB-HMXB$^{\ast}$ & AIC(?)$^{\dagger}$ \\
\colrule
$0 < B < 10^{12}$ & "normal" SN & spherical, weak short GRB & UV-X flash \\
$B \sim 10^{13}$ & bipolar SN & bipolar,strong short GRB & bipolar UV-X flash \\
$ B \geq 10^{14}$ & ? & jet-like, weak short GRB & jet-like UV-X flash \\
$B \gg 10^{15-16}$ & -- & -no SQM formation- & -- \\ \botrule
\end{tabular}}
\end{table}

\noindent $\ast$ only if $NM \rightarrow SQM$ conversion is
sometimes suppressed when a NS is formed.

\noindent $\dagger$ upper limit to the rate $\sim 10^{-4} yr^{-1}
galaxy^{-1}$ needs to be revised if SQM burning occurs modifying
nucleosynthetic yields.

\section{Detection of SQM}

Although the present work has been intentionally biased towards
the astrophysical aspects of SQM, the importance of laboratory
experiments can not be overstated. The main well-known
disadvantage of heavy ion collision experiments for the search of
SQM production is clearly the high-temperature environment, which
tends to destabilize small SQM chunks (strangelets) and render
their production difficult. In addition, even if produced, the
''dirty'' high particle multiplicity environments makes the
identification of strangelets a complicated business. However, the
recent claims \cite{CERN} for QGP production already before the
RHIC runs (see Ref.~\refcite{RHIC} and the literature therein) and
the monotonic trend of the observables in support of the former
should revive the interest in strangelet physics, given the the
precondition of QGP formation seem likely.

An alternative for the direct detection relies on the improvement
of detector analysis in a variety of cosmic rays and related
experiments. For example, the prospects for the space-based
spectrometer AMS are encouraging \cite{AMS} since exciting claims
about nuggets crossing the earth have been issued \cite{Sys} and
merit a closer examination.

Returning to the stellar astrophysics, SQM may be ''detected'' in
a binary system when lines observed from one member of the source
allows in some cases a determination of the mass $M$ and redshift
$z$, and thus a knowledge of the radius $R$ (the same idea has
been used in the '20s to determine structure of WDs). Previous
work using the redshift of spectral lines involved mainly isolated
NSs, like the exciting announcement of Cottam, Paerels and
M\'endez \cite{CPM} of a redshift $z = 0.35$ detected in the
bursting source EXO 0748-676 spectrum. Contrary to the naivest
analysis, we must stress that this determination does not exclude
at all a SS as a candidate as claimed in the latter work, because
we can not still pin down the type of viable SS models (see
Ref.~\refcite{Xu}). It is probably more accurate to state that if
the source redshift determination is correct and a SS model is
constructed for it, then SQM would produce very compact sources
for very low values of the mass only (see Ref.~\refcite{BH0}).

The most promising advances in the detection  of SQM are those
related to the observation of compact star structure and cooling.
The goal is to determine whether NS are composed of ordinary
beta-stable nuclear matter or contain QCD phases in their
interior. This has a direct observable impact on the global static
properties of the star (such as the mass-radius relationship) and
in the short and long term cooling history of these objects.
Briefly, objects with QCD phases inside them tend to have smaller
radius and to cool faster than beta-stable nuclear matter objects
of the same mass. Some mass and radius determinations made up to
date, have yielded values around 1 solar mass and $\sim 7 km$
(e.g. the sources Her X-1 \cite{Dey} and  SAX J1808.4-3658
\cite{Li}). Other sources offer better determinations but are
still controversial (RXJ 1856-37 \cite{Pons} and  EXO 0748-676
\cite{CPM} mentioned above, among a few others). There are some
tentative indications that the value of the surface temperature of
some objects (e.g. 3C58, Vela, Geminga) cannot be completely
understood in terms of the standard cooling theory (see
Ref.~\refcite{Tsu} and references therein). Therefore, it has been
argued that these objects cannot be normal neutron stars and
should have exotic phases in their interior. Many unresolved
questions are relevant in connection with this picture, since they
are expected to have strong impact in the observational output
(see Ref.~\refcite{LH1} for discussion). These include the
appearance  of superfluidity in dense stellar matter, a deeper
understanding of the QCD phase diagram in the high density regime,
the improvement of phenomenological models for strong interactions
at finite density, a comprehension of the dynamics of a possible
hadron-quark phase transition in stellar conditions and an
evaluation of the related consequences and signatures in each
specific astrophysical context.

\section{Surprises on the microscopic state of SQM}

\begin{figure}
\includegraphics[angle=-90,width=12cm,clip]{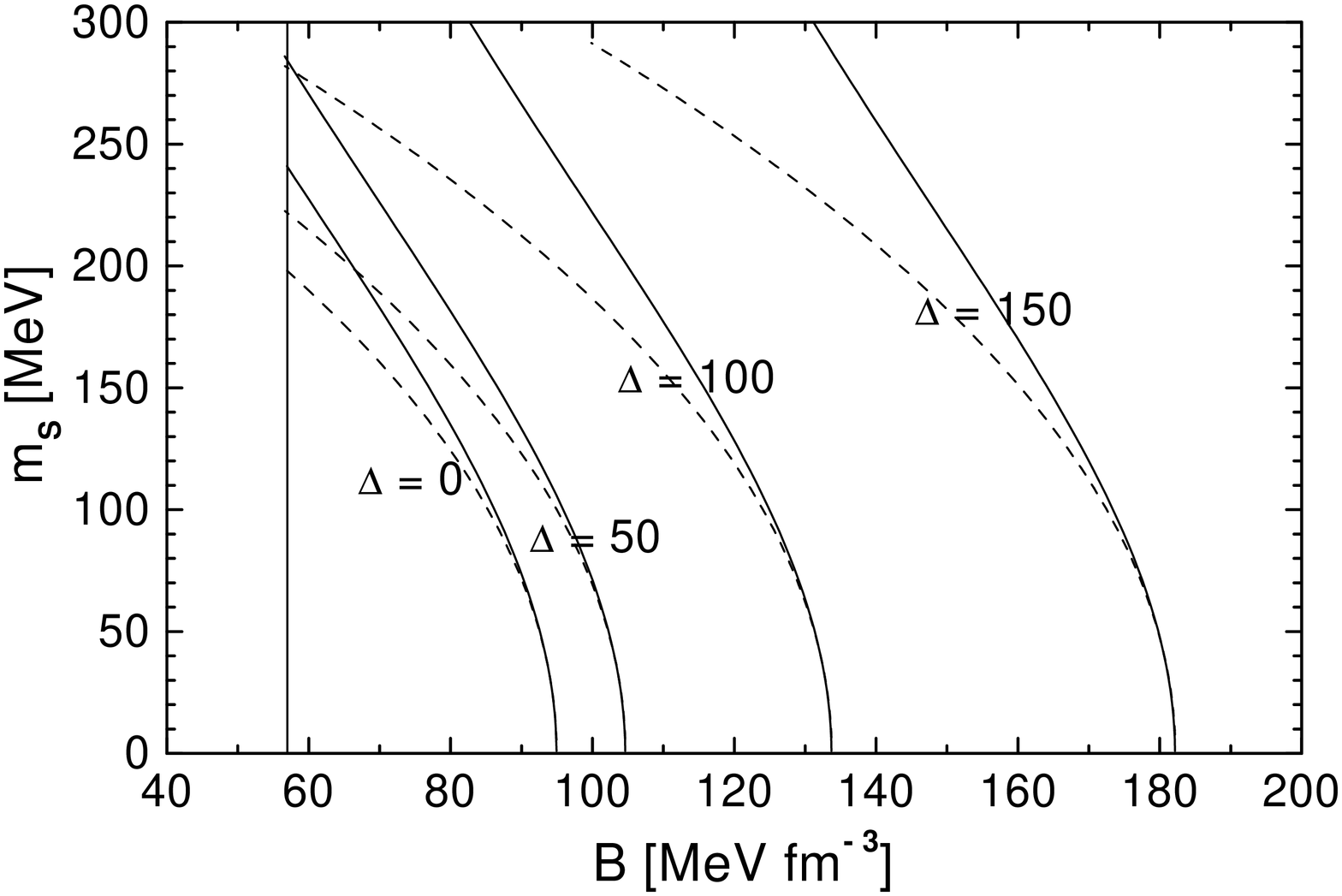}
\caption{The window of stability of CFL strange matter. Room for
an absolutely stable CFL state lies between the boundaries. Each
window refers to a given value of the gap $\Delta$ as in
indicated. The regions are much wider than the ones found for
SQM$^{12}$. The dashed lines display the approximated result to
order $m_{s}^{2}$, as discussed in Ref. 51.}
\end{figure}

The very basic picture of SQM as a Fermi liquid confined by some
parametric ingredient (generally put by hand) leaves lots of room
for improvements and questioning. Asymptotic freedom is expected
to drive the system beyond a certain density that can not be
reliably determined today. It is quite natural to expect that a
more complex picture will emerge when a closer look to the system
is taken. For example, while {\it correlations} between quarks are
certainly expected in the hadronic phase and immediately above the
deconfinement point \cite{Fod}, the role of two-quark states known
as {\it diquarks} did not receive much attention until a few
papers \cite{DS} pointed out how useful they may be in the
description of dense stellar matter. Simple models were
constructed in the spirit of a partial bosonization of the quark
phase, and the corresponding stellar sequences studied. Later, a
revival of the study of QCD pairing leaded to identify some
interesting possibilities (two-flavor superconductor -2SC- and
color-flavor locked -CFL- state) that gave a whole new twist to
the description of dense matter \cite{Al}. It is currently
believed that the CFL state is energetically favorable over the
2SC and therefore may comprise a large part of stellar interiors.
An important point related to our discussion is the recognition of
the possible role of pairing in the absolute stability of matter.
Detailed calculations performed for strangelets \cite{JM} and bulk
strange matter \cite{LH2} showed the substantial enlargement of
the stability window (Fig. 3) due to the action of the gap energy
$\Delta$ when compared with the gapless state. This "color-flavor
locked strange matter" allowed the calculation of self-bound stars
constructed with parametric equations of state in
Refs.\cite{LH3,AR}.

A discussion of the effects of a sudden release of this extra
energy has been presented by Hong, Hsu and Sannino \cite{Hsu} and
subsequently criticized by Blaschke et al. \cite{Dav1}. The main
issue whether the pairing, acting immediately after deconfinement,
will release the energy in a "useful" form (i.e., to power a
supernova), a question which is also related to the timescale of
this event. In any case neutrinos would be produced and their
(sudden) release add to the energy budget, as advocated in
Ref.~\refcite{Dav2}.

Even when acknowledged that the physical description of the
CFL state is still crude, it is important to have in mind that
substantial surprises may be hidden in the high-density region of
the QCD phase diagram, with all their implications for
astrophysical studies.

\section{Types of published contributions to SQM and their challenges}

Given the different situations in which SQM may be important, it
may be asked how does the literature published on SQM reflect that
trends. It can be said that the "birth" (prior to 1984) and "early
infancy" ($\geq \,  1984$) papers already shown the general spirit
of the later work in the field, which in essence has not changed
substantially but altered somewhat their proportional weights.
Generally speaking, and following  these pioneer efforts,
follow-up papers may be classified according to their content in
a) papers trying to address the stability itself (generally within
a model calculation) and SQM physical properties; b) papers
devoted to the search of SQM in laboratory and related
environments c) papers dealing with astrophysical/cosmological
SQM, including observations which may help to prove/disprove the
SQM presence far away from the laboratories. All three things are
quite difficult to argue, but for different reasons. The class a)
needs to convince the community that the models/calculations are
as accurate as $\sim 1 \, \%$, since the difference $(m_{n} \, -
\, {E/n})_{P=0}$ is not expected to be larger than few tens of
$MeV$s. Class b) works by ever closing windows in the parameter
space, although loopholes in the experimental settings or
measurements may be found from time to time (and in fact they are,
an event that shifts the excluded region in a significative
manner). Finally, class c) suffer from both the uncertainties of
the astronomical measurements themselves and also from the danger
that colleagues astronomers devise an interpretation at least as
solid as the SQM-based one but invoking more conventional physics
(a task which generally takes a short time after the observation
being reported). In other words, and in spite of serious attempts
to find the Holy Grail of SQM physics, we still lack of convincing
evidence of its reality, a statement that also applies to any
clear-cut astrophysical/cosmological crucial test hitherto
proposed. One positive tendency, related to the degree of maturity
of the field, is the increasing number of realistic models that
going well beyond the basic SQM properties attempt a fine
comparison and sometimes a synthesis to fit in the observed
phenomena. A clear example of this attempts (not fully successful
as yet) is the efforts to understand the room for strange stars in
the compact object zoo \cite{Frido}. As previously mentioned,
direct search in space is an exciting perspective because large
flux estimates have been obtained \cite{Jes1}.

\section{Statistics, facts and outlook}

\begin{figure}
\includegraphics[angle=-90,width=12cm,clip]{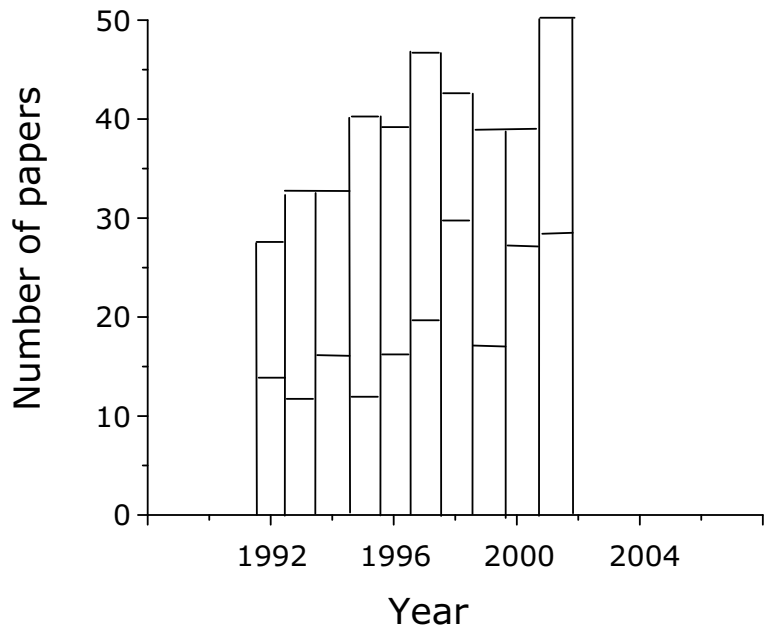}
\caption{The number of SQM papers 1992-2001. The marks  in each
bar shows the number of type c) papers dealing with
astrophysics/cosmology. It should be remarked that the selection
has not been extremely rigorous to avoid the omission of some
contributions helpful for those interested in the full listing.
This means that no distinction has been made between,say,
proceeding and journal papers (often containing the same results)
for the sake of completeness. The sources (SPIRES, Web of Science,
ADS and a few others) are well-known extensively used databases,
and the completeness of the sample is likely to be $> 95 \%$,
which allows to draw safe conclusions and locate the searched
material in most cases. Any omitted contributions, specially from
less visible sources, may be communicated to the authors to be
included in the list.}
\end{figure}

Can we visualize any definite trend in the literature and forecast
somewhat the future of SQM ? The answer may depend on what exactly
is focused. The first interesting point to notice is that the
research papers in SQM have grown beyond an expected vegetative
rate, perhaps reflecting the growth in the number of people
involved in sciences overall and the attractive of a ``magnet''
field. The absolute number of papers (390) directly dealing with
SQM in the last decade (1992-2001) is shown in Fig. 4. There is a
smooth continuous growth of $\sim 70 \%$ in this number of papers
respect to the previous decade, to be compared with the absolute
number  (223) in the previous Aarhus compilation comprising the
period 1969-1991. The lower bars in Fig. 4 correspond to the
papers classified as c), which have grown from about $1/3$ of the
total at the beginning of the decade to more than one-half in the
latest years (with a peak of $\sim \, 2/3$ in year 2000). However,
it is interesting to note that  class c) papers already comprised
a $\sim 60 \%$ of the total at the time of the Aarhus compilation.
This means that after some years in which the
astrophysical/cosmological papers lost their leadership, the
situation bounced back to the pre-1991 average value. On na\"ive
grounds, an outsider physicist just aware of the status of SQM
would have expected the predominance of type b) based on the
referred lack of experimental/observational confirmation of the
SQM existence \cite{Klin}. To be sure, class c) papers do address
to some extent this issue indirectly, but the importance of the
experimental evidence can never be overstated. Our impression is
that, unless evidence for SQM is available along the first decade
of the XXI century, there would be a decline (not just a slowdown
of the growth rate). It is however quite clear that announcements
of data which is difficult to interpret within conventional
hadronic models (see previous sections) may, if sustained, give a
substantial impulse to SQM, taking the place of the laboratory
data to a large extent. Indeed, it is not impossible that physical
features preclude the laboratory detection for indefinite time but
astrophysical/cosmological evidence becomes available. It is not
easy to foresee what kind of trend would this hypothetic situation
(i.e. lack of laboratory evidence but definite astrophysical
hints) could create. The case would resemble, for instance, the
theory of matter inside the white dwarfs, never produced in
laboratory but widely accepted by all astrophysicists.

As a concluding remark we should add that SQM is to some of us a fundamental
issue too important to be dismissed (or confirmed) at present. Even if proved to
be wrong, the SQM hypothesis has stimulated a deep analysis by theoreticians and
experimentalists as well that goes much beyond the very limited subject.
This facts together with several unique features present in the development of the
field has provided, and would likely provide, an excellent opportunity to
learn about nature and the way knowledge is constructed in a vivid and fruitful fashion.

\section{Acknowledgements}

Along several years of SQM research many people contributed to clarify and
explain features of the physics and astrophysics of SQM to the authors, and
provided guidance in many respects. Among them we wish to acknowledge O.G.
Benvenuto, H. Vucetich,  J.A. de Freitas Pacheco and I. Bombaci, colleagues and friends.
The authors wish to thank the S\~ao Paulo State Agency FAPESP for
financial support through grants and fellowships, and the partial support
of the CNPq (Brazil).

\end{document}